\documentclass[twocolumn,aps,prd,showpacs,nofootinbib,superscriptaddress,preprintnumbers,floatfix,10pt]{revtex4-2}

\usepackage[english]{babel} 
\usepackage[utf8]{inputenc}
\usepackage[T1]{fontenc}
\usepackage{lmodern}

\usepackage{amsmath}
\usepackage{amsfonts}
\usepackage{amssymb}
\usepackage{slashed}
\usepackage{bbm}

\usepackage{graphicx}
\usepackage{color}
\usepackage{multirow}

\usepackage{hyperref}

\newcommand*{\no}{\noindent}
\newcommand*{\bea}{\begin{eqnarray}}
\newcommand*{\eea}{\end{eqnarray}}
\newcommand*{\be}{\begin{equation}}
\newcommand*{\ee}{\end{equation}}

\newcommand*{\pref}[1]{(\ref{#1})}

\newcommand*{\prefr}[2]{(\ref{#1}-\ref{#2})} 
\newcommand*{\nn}{\nonumber}

\newcommand{\bma}{\begin{pmatrix}}
\newcommand{\ema}{\end{pmatrix}}

\newcommand*{\la}{\left\langle}
\newcommand*{\ra}{\right\rangle}

\begin{document}

\title{Restoring the Bloch-Nordsieck theorem in the electroweak sector of the standard model}

\author{Axel Maas}
\email{axel.maas@uni-graz.at}
\author{Franziska Reiner}
\email{franziska.reiner@edu.uni-graz.at}
\affiliation{Institute of Physics, NAWI Graz, University of Graz, Universit\"atsplatz 5, A-8010 Graz, Austria}

\begin{abstract}
The electroweak gauge symmetry cannot be broken in a literal sense due to Elitzur's theorem. Thus, asymptotic states need to be manifestly and non-perturbatively gauge-invariant with respect to the electroweak symmetry. To take this suitably into account perturbation theory augmented by the Fr\"ohlich-Morchio-Strocchi mechanism can be used. We show that this restores the Bloch-Nordsieck theorem in electroweak processes in the standard model. This has potentially substantial impact at, e.\ g., future lepton colliders, but has only negligible effects at lower energies. We also demonstrate an alternative implementation using PDFs, which allows an approach with manifest electroweak Bloch-Nordsieck theorem also at hadron colliders.
\end{abstract}

\maketitle

\section{Introduction}

Gauge symmetries can never be broken spontaneously due to Elitzur's theorem \cite{Elitzur:1975im,Frohlich:1980gj,Frohlich:1981yi}. As a consequence, the electroweak gauge symmetry is actually intact, and its apparent breaking is only a consequence of a gauge choice \cite{Lee:1974zg}. In fact, it is possible to define gauges in which it remains unbroken \cite{Lee:1974zg,Maas:2012ct}, at the expense of the applicability of perturbation theory \cite{Lee:1974zg}. Hence, as has been realized early on \cite{Banks:1979fi,Fradkin:1978dv,Osterwalder:1977pc,Frohlich:1980gj}, physical, observable states need to be manifestly non-perturbatively gauge-invariant, rather than only perturbatively BRST invariant\footnote{The reason for BRST invariance to be insufficient is the Gribov-Singer ambiguity \cite{Gribov:1977wm,Singer:1978dk}, which explicitly invalidates perturbative BRST \cite{Fujikawa:1982ss}, see \cite{Maas:2017wzi} for a review.}, also with respect to the electroweak symmetry \cite{Frohlich:1980gj,Frohlich:1981yi}. This requires to consider composite states rather than elementary ones as asymptotic in states and out states, very much like hadrons in QCD.

At first sight this seems to be in contradiction to the highly successful description of experiments in terms of the elementary states in perturbation theory. This is explained by the Fr\"ohlich-Morchio-Strocchi (FMS) mechanism \cite{Frohlich:1980gj,Frohlich:1981yi}. It shows that in the gauges usually employed in perturbation theory many quantities are dominated, up to corrections suppressed by powers of the ratio of the relevant energy scale to the Higgs vacuum expectation value, by their usual perturbative expression. Augmenting perturbation theory with the FMS mechanism appears to allow to take the remaining deviations manifestly into account analytically\footnote{Similar considerations have been followed already in \cite{Jegerlehner:1985ch,Jegerlehner:1984ia}.} \cite{Maas:2020kda,Dudal:2020uwb}. This approach has been supported by lattice calculations, and is reviewed in \cite{Maas:2017wzi}. 

In section \ref{s:lepton} we will briefly review the mechanism and then apply it to fermionic 2-to-2 processes at lepton colliders. We will extend here \cite{Egger:2017tkd} to formulate the full FMS expression of the relevant matrix elements. We thereby show how systematically at low energies the ordinary results will appear, while at the same time cancellations ensure at high energies that the violation encountered in standard perturbation theory \cite{Ciafaloni:2000df,Ciafaloni:2000rp} of the Bloch-Nordsieck theorem (BNT) \cite{Bloch:1937pw,Agarwal:2021ais} is absent. We will briefly discuss the implications for future lepton colliders in a sample processes.

The same is true, in a more involved way, for hadronic initial states \cite{Egger:2017tkd,Maas:2017wzi}. The strongly interacting nature does not allow a purely perturbative approach as in the lepton case. Rather, we expand the usual PDF language developed for $pp$ collisions in a gauge-invariant way in section \ref{s:hadron}, making the BNT manifestly maintained. This yields a few surprising features of its own as to what the actual PDFs represent at high energies, and what kind of sum rules they need to obey.

We note that similar considerations will also apply to violations \cite{Ciafaloni:2022kil} of the Kinoshita-Lee-Naunberg theorem (KLNT) \cite{Kinoshita:1962ur,Lee:1964is,Agarwal:2021ais}. Moreover, beyond leading order, it is in general insufficient to satisfy only the BNT to achieve infrared-safe observables, but the KLNT needs to be satisfied instead \cite{Agarwal:2021ais,Doria:1980ak,Andrasi:1980qw}. While this is beyond the scope of the present work, we briefly comment on consequences and necessary steps in section \ref{s:kln}.

We wrap up the discussion in section \ref{s:sum} with a number of possible further steps. Some preliminary results have been made available in \cite{Reiner:2021bol}.

\section{Lepton scattering}\label{s:lepton}

\subsection{Weakly gauge-invariant leptons}

We consider in the following the usual standard model Lagrangian, supplemented for simplicity by right-handed neutrinos, following \cite{Frohlich:1980gj,Frohlich:1981yi,Egger:2017tkd,Afferrante:2020fhd}. The existence of right-handed neutrinos is not essential for this work, but they simplify considerably the technicalities. Also, they allow for a very straightforward transfer of results between quarks and leptons.

As emphasized in the introduction, field-theoretical arguments highlight that the weak gauge symmetry is never spontaneously broken, and what is usually called the Brout-Englert-Higgs (BEH) effect is a particularly suited choice of gauge, which allows for a technically advantegous treatment of the dynamics \cite{Elitzur:1975im,Banks:1979fi,Fradkin:1978dv,Osterwalder:1977pc,Frohlich:1980gj,Frohlich:1981yi,Lee:1974zg}. A full discussion of the details and background can be found in the review \cite{Maas:2017wzi}.

The most important consequence is that weakly gauge-dependent states, especially the Higgs, the $W$ and $Z$ bosons, and all left-handed fermions cannot be physical. Rather, just like in QCD, only manifestly gauge-invariant composite states can be physical asymptotic states \cite{Banks:1979fi,Frohlich:1980gj,Frohlich:1981yi}. We will concentrate here on the fermions, and refer for the bosons again to the review \cite{Maas:2017wzi}. In this context, it is useful to write the complex doublet Higgs field $\phi$ before separating off the vacuum expectation value (vev) as
\be
 X = \begin{pmatrix} \phi_{2}^{*} & \phi_{1} \\ -\phi_{1}^{*} & \phi_{2} \end{pmatrix}.
\label{eq:higgs_matrix}
\ee
\no and thus the standard Higgs fluctuation mode as well as the three would-be Goldstone bosons are still contained in $X$.

Fermions come now in two varieties. One are the left-handed Weyl spinors $\psi^L$, which are gauged under the weak interaction in the fundamental representation. I.\ e., they form doublets \be
\psi^L = \bma \nu^L \cr e^L\ema\label{ldoublet}
\ee
\no of which there are six, three for the lepton generations and three for the quark generations. In addition, there are twelve right-handed Weyl spinors, representing the six ungauged leptons and quarks of which always two can be paired, e.\ g.\ $\nu^R$ and $e^R$, as an ungauged set corresponding to the left-handed Weyl doublet. Besides the strong interactions separating quarks and leptons the other differences arise due to the Yukawa couplings, hypercharge, and CKM/PMNS values for the, in total, 18 fermion fields.

At vanishing hypercharge and Yukawa couplings, there are several symmetries in the weak sector of the standard model. First, there is the local weak gauge symmetry $\mathrm{SU(2)_{w}}$. Second, there is a global $\mathrm{SU(6)_{Rf}}$ flavor symmetry of the right-handed Weyl fermions and an SU(3) left-handed generation symmetry. These are present for quarks and leptons separately. Finally, there is a less obvious global $\mathrm{SU(2)_{c}}$ symmetry which acts only on the scalar doublet as a right-multiplication on $X$. Switching on Yukawa couplings breaks these symmetries to the familiar pattern of lepton and quark number, and a U(1) subgroup of $\mathrm{SU(2)_{c}}$. Gauging the latter adds hypercharge.

Ignoring for a moment the BEH effect, there are for each generation and for quarks and leptons separately four physical fermionic states in the theory, which can be grouped into two chiral doublets. The first two states are the flavor doublet of right-handed Weyl fermions $\chi^{R}$. One of them is the right-handed charged lepton, $\chi^{R}_{2} = e^{R}$ and the other the right-handed neutrino $\chi^{R}_{1} = \nu^{R}$. The other physical doublet is a gauge-invariant, left-handed composite Weyl field, $\Psi^L=X^\dag\psi^L$ \cite{Frohlich:1980gj,Frohlich:1981yi,Egger:2017tkd}, which is a singlet with respect to the non-Abelian gauge group but carries a global $\mathrm{SU(2)_{c}}$ charge. The two components of this doublet will be identified with the left-handed electron and the left-handed neutrino below.

In case non-zero Yukawa couplings break the global $\mathrm{SU(2)_{c}}$ symmetry and the flavor symmetry $\mathrm{SU(2)_{Rf}}$ to the diagonal subgroup $\mathrm{SU(2)_{f}}$. The composite state and the right-handed fermion doublet transform in the same way under $\mathrm{SU(2)_{f}}$. In this way it appears as if in the physical spectrum the diagonal subgroup acts as an effective flavor symmetry for both the left-handed sector and right-handed sector simultaneously. Note that the two gauge-dependent components of the elementary left-handed Weyl fermion $\psi^{L}$ do not transform under $\mathrm{SU(2)_{f}}$, but can be transformed into each other via a gauge transformation and can therefore not be associated with physically observable particles or flavors. 

To connect to the usual perturbative picture, switch on the BEH effect in a 't Hooft gauge. The Higgs field can then be split into its vacuum fluctuations $\eta$ and its vev $v$. Using the gauge freedom, we conventionally chose the vacuum expectation value to be in the real 2 direction, $\langle \phi_{i} \rangle = v \delta_{i2}/\sqrt{2}$. At tree-level, this yields the customary result that the gauged and ungauged Weyl spinors can be combined into two Dirac spinors, each with a mass given by $m_f = y_f v/\sqrt{2}$, forming the usual leptons \cite{Bohm:2001yx}.

However, the left-handed leptons are thus defined in a gauge-fixed way. The decisive step to relate them to the gauge-invariant $\Psi^L$ is to apply the FMS mechanism \cite{Frohlich:1980gj,Frohlich:1981yi}. Consider the physical left-handed fermion\footnote{This construction automatically ensures the correct assignment of hypercharge, and by extension of electric charge, to all bound states \cite{Maas:2017wzi}. It will therefore not be considered further in the following.} \cite{Frohlich:1980gj,Frohlich:1981yi},
\be
\Psi^{L} = X^\dag\psi^L=\left(\frac{v}{\sqrt{2}}\mathbbm{1}+ \eta\right)\psi^L
= \frac{v}{\sqrt{2}} \bma \psi_1^L\\ \psi_2^L\ema+\eta\psi^L\label{expferm},
\ee
\no where the matrix-valued $\eta$ contains the usual fluctuation field identified with the elementary Higgs boson and the Goldstone fields in the same manner as $X$ contains $\phi$ in \pref{eq:higgs_matrix}. To leading order in $v$, the first term, the field $\Psi^L$ reduces to the elementary left-handed fermions, which are thus the FMS-dominant constituents of the bound state $\Psi^L$. Of course, only the total sum in \pref{expferm} is gauge-invariant, and the leading order alone is not.

When now forming a propagator it follows
\be
\la \Psi_{f_{1}}^L(p) \bar{\Psi}_{f_{2}}^L(-p)\ra = \frac{v^2}{2} \la \psi_{f_1}^L(p) \bar{\psi}^L_{f_2}(-p) \ra+{\cal O}\left(\frac{|p|}{v}\right)\label{expferm2}.
\ee
\no Thus, to all orders in perturbation theory and to leading order in $|p|/v$ the propagator of the physical, composite fermion state is given by the gauge-dependent elementary ones, i.e., the propagators of the left-handed charged lepton and neutrino. Especially, the poles and thus the masses and widths coincide. This was shown to all orders in perturbation theory for the Higgs bound-state--elementary-state duality \cite{Maas:2020kda,Dudal:2020uwb} and appears to generalize straightforwardly to all other standard model particles \cite{Maas:2020kda,Dudal:2020uwb}. In particular, propagators of the type \pref{expferm2} indeed are necessarily also off-shell gauge-invariant when including the finite polynomial in $v$ in each order in perturbation theory. Only at $|p|\gg v$ they start to appreciably deviate from the perturbative result at the same order \cite{Maas:2020kda,Dudal:2020uwb}.

In this way, the gauge-invariant Dirac spinor $(\Psi_f^L\;\chi_f^R)^\mathrm{T}$ describes the physical neutrinos ($f=1$) and charged leptons ($f=2$) with the same properties as the usual gauge-dependent ones of perturbation theory \cite{Frohlich:1980gj,Frohlich:1981yi,Maas:2017wzi}. This can be extended to include the hypercharge sector \cite{Maas:2017wzi,Egger:2017tkd}, and quarks work completely analogously apart from complications due to the strong interactions \cite{Maas:2017wzi,Egger:2017tkd}.

All lattice results \cite{Maas:2017wzi,Maas:2018ska,Maas:2020kda,Afferrante:2020fhd,Jenny:2022atm} as well as the fact that perturbation theory is a very good approximation so far to experiments imply that the corrections to perturbation theory are indeed small, at least at energies at or below the electroweak scale. But the important bottom line is that the physical left-handed leptons in the standard model are actually bound states of the elementary ones and the Higgs field \cite{Frohlich:1980gj,Frohlich:1981yi}.

\subsection{Scattering}

In fact, the subleading contribution in \pref{expferm2} are not zero, and are thus, in principle, detectable. In vector-boson scattering and form factors these have indeed been isolated in a reduced standard-model setup on the lattice \cite{Jenny:2022atm,Maas:2018ska} and for the off-shell properties of the Higgs and the vector bosons analytically \cite{Maas:2020kda,Dudal:2020uwb}. But it is far from straightforward to observe them experimentally, yet. However, the kinematic suppression in \pref{expferm2} and \cite{Maas:2020kda,Dudal:2020uwb,MPS:unpublished,Fernbach:2020tpa,Egger:2017tkd} suggests that this will change if the energy scale is substantially above the electroweak scale.

We therefore extend here the ad-hoc approximations in exclusive fermion scattering of \cite{Egger:2017tkd}. Consider a polarized cross-section of $\bar{f}f\to\bar{F}F$, where the initial fermions $f$ and final fermions $F$ are distinct. If the energy scales are large enough compared to the fermion masses, the scattering can be considered polarized, i.\ e.\ each of the $f$ and $F$ are either completely left-handed or right-handed. In the purely right-handed case both initial state and final state are entirely made-up from weak gauge-singlets, and nothing will change. Thus, concentrate on at least one of them being left-handed.

The relevant matrix elements are\footnote{Note that also the LSZ formalism needs to be suitably augmented for external bound states, which is straightforward \cite{Bohm:2001yx,Weinberg:1995mt}. What is not straightforward is how this interacts with the FMS expansion beyond the leading term \pref{loxs} below. This will be presented elsewhere \cite{MPS:unpublished}. However, as long as the Higgs contents in the bound state is small compared to the one of the fermionic component, this will only be a minor effect \cite{Egger:2017tkd}.}
\bea
&&\la\bar{f}^R f^L \bar{F}^R F^L\ra\label{fbs1}\\
&&\la\bar{f}^L f^L \bar{F}^L F^L\ra\label{fbs2},
\eea
\no and $L$-$R$ permutations in \pref{fbs1}. Following \pref{expferm}, this will reduce to the usual perturbative expression to all orders in the coupling and to leading-order in the FMS expansion \cite{Egger:2017tkd}. By dimensional analysis, the higher orders in the FMS expansion will be suppressed by $s/v^2$, where $s$ is the relevant largest energy scale, e.\ g.\ the center-of-mass energy at tree-level. Thus, to have something interesting happening, $s\gtrsim v^2$ is needed.

While an explicit calculation of first-order corrections is possible along the lines of \cite{Maas:2020kda,Dudal:2020uwb} and is under way \cite{MPV:unpublished}, there is a much more generic effect if $s\gg v^2$, and thus $s\gg m_{W/Z}^2$. This would be the situation at a TeV-scale lepton collider. In such a case in the standard perturbative approach the fact that the external left-handed fermions are gauge-dependent makes itself felt by a violation \cite{Ciafaloni:2000rp,Bauer:2018xag,Manohar:2018kfx} of the BNT \cite{Bloch:1937pw}. Because the initial state and final state are build from gauge-non-singlets, the usual summation over the full gauge multiplet needed for cancellations of infrared divergencies in the BNT does not occur, and they remain in form of double Sudakov logarithms $\sim\ln^2(s/m_W^2)$ \cite{Ciafaloni:2000rp} from emissions of (electro)weak gauge bosons \cite{Fadin:1999bq,Ciafaloni:1999ub}, cut off only by the weak mass scale $m_W$. At a TeV lepton collider the effect is of the same order as strong corrections, i.\ e.\ a couple of percent \cite{Ciafaloni:2000rp}, and would grow with larger energies.

By its very construction, it is clear that a scattering process only involving \prefr{fbs1}{fbs2} necessarily respects the BNT in general. At the investigated very large energies, the BEH effects is irrelevant, and all masses can be neglected. Since the gauge multiplets are now complete, the proof that the BNT holds is completely analogous to the one in QCD \cite{Bohm:2001yx}, except for the fact that the lower cutoff is fixed by the weak mass scale: Just as in QCD\footnote{Neglecting the vev, this can also be shown using a coherent state approach completely analougsly to the QCD case, except for the presence of the Higgs self-interactions. However, if the vev is neglected, there is only an interaction of four Higgs possible, which is thus as irrelevant as the four-gauge boson interaction.} any quark color can contribute, and thus need to be summed over, all weak charges of the doublet \pref{ldoublet} in the bound state \pref{expferm} can equally likely contribute, and thus need to be summed over. Or stated otherwise, by having the bound state \pref{expferm} rather than the weak doublet \pref{ldoublet} involved as external states in \prefr{fbs1}{fbs2}, the process becomes {\it {\bf inclusive} with respect to the gauge doublet \pref{ldoublet}}, with the Higgs components of the bound state playing the same spectator role as the remaining spectator quarks of a proton in QCD. It is only {\it {\bf exclusive} at the level of the global SU(2)$_\text{c}$ components of the bound state \pref{expferm}}.

While this statement is necessarily true, it is nevertheless very interesting how this works out in practice. And especially why this process does not deviate at low, i.\ e.\ LEP2, energies, appreciably from the ordinary result involving elementary particles as asymptotic states. Since the violation of the BNT has at leading-double-log order only implications for the scattering for two-left-handed particles \cite{Ciafaloni:2000rp}, we will concentrate on this case. Consider for concreteness the process $\bar{l}l\to\bar{q}q$, as here the violation has been explicitly worked out in \cite{Ciafaloni:2000rp}, and thus here only the changes need to be addressed.

The corresponding matrix element is then
\begin{equation}
\left\langle \bar{\Psi}^L_2\Psi^L_2 \Bar{Q}^L_2 Q^L_2\right\rangle=\left\langle\underbrace{\bar{\psi}^L_i X_{i2}}\underbrace{X^\dagger_{2j}\psi^L_j}\underbrace{\bar{q}^L_k X_{k2}}\underbrace{X^\dagger_{2l}q^L_l}\right\rangle\label{me}
\end{equation}
where $Q$ is the composite state for the left-handed top-type quark and bottom-type quark constructed analogously to \pref{expferm} from  $q^L=(b^L,t^L)^T$. For concreteness, we consider the $\bar{e}e\to\bar{c}c$ process, and select the custodial indices accordingly to be 2. Underbraces have been used to identify combinations, which make up composite operators. This implies that all composite operators in the initial state and final state involve both members of the weak fermion doublet \pref{expferm}

First, we investigate what happens when performing the FMS mechanism in leading order in $s/v$. In this case, $X=v\bf{1}$, and the expression collapses as \cite{Egger:2017tkd}
\begin{equation}
\left\langle \bar{\Psi}^L_2\Psi^L_2 \Bar{Q}^L_2 Q^L_2\right\rangle=v^4\left\langle \bar{e}^L e^L \bar{c}^L c^L\right\rangle+\mathcal{O}\left(\frac{s}{v}\right)\label{loxs},
\end{equation}
where $s$ is again the dominating energy scale, at tree-level the center-of-mass energy \cite{Egger:2017tkd}. Thus, to this order, we recover the usual expression for the matrix element. This shows how the FMS mechanism recovers the usual result, if the finite number of other terms in the FMS expansion can be neglected. Thus, at LEP2 energies this should be a fairly good approximation. However, the BNT violation also plays no role at these energies as $s\sim m_W^2$ \cite{Ciafaloni:2000rp,Bauer:2018xag,Manohar:2018kfx}, so it is no contradiction that the violation appears to be recovered. 

This can therefore be no longer be the case if $s\gg m_W^2$, like at future lepton colliders. Assuming a fully {\it exclusive measurement of the final state SU(2)$_\text{c}$ quantum numbers}, however, it is safe to assume that the Higgs fields in \pref{me} act primarily as spectators to the FMS-dominant constituent. This is also supported by investigations of the substructure of the bound state \pref{expferm} \cite{Egger:2017tkd,Afferrante:2020fhd}. This is akin to the situation in hadron scattering where the other partons are also spectators. Of course, the Higgs are then not leaving as debris, but will be needed to construct again weakly gauge-invariant final states, dressing again the elementary states \cite{Egger:2017tkd}. In that sense, the process is different than in QCD, where the dressing is obtained in fragmentation from soft gluons and light quarks. Due to the large masses and mass defects, this does not happen at this level in an electroweak process. However, at sufficiently high energies, electroweak fragmentation of elementary weak particles into bound states like \pref{expferm} in the final state can be expected to also happen very similarly to QCD.

As the process is then {\it inclusive with respect to the weak doublet \pref{ldoublet}}, it is thus necessary to sum the cross-sections of the possible pairings. Ignoring for a moment the final state, this implies that the total cross section is given by\footnote{In fact, the same result would be obtained when using a PDF description for the bound state \pref{expferm}, by setting $f_e=f_\nu=\delta(x)$ and all Higgs PDFs to zero in \pref{pdf}.}
\begin{equation}
\sigma_{\bar{\Psi}_L^2\Psi_L^2\to X}\sim\sigma_{\bar{l}_L l_L\to X}+\sigma_{\bar{l}_L \nu_L\to X}+\sigma_{\bar{\nu}_L l_L\to X}+\sigma_{\bar{\nu}_L \nu_L\to X}\label{tsx}.
\end{equation}
\no This neglects the Higgs spectator interactions, and will therefore have corrections at higher orders. The fact that the Higgs behave as spectators has been suppressed in the notation, just as with hadron-hadron collisions. Nonetheless, because the net zero weak charge of the bound states can be arbitrarily split between the Higgs spectator and the elementary lepton field, a summation is necessary over all possible weak charge states, as given in \pref{tsx}. This is again in complete analogy to color in QCD.

However, by relegating the Higgs to spectators, only the fermionic components of the bound state \pref{expferm} can now interact, especially also the left-handed neutrinos. This is, because the experimentally prepared initial states are SU(2)$_\text{c}$ eigenstates and not unphysical gauge eigenstates. Just like all colors are contained for every valence quark in the proton, now all weak charges, and thus left-handed electrons and left-handed neutrinos both, are contained in the initial SU(2)$_\text{c}$ eigenstate, and thus need to be summed over\footnote{Note that the usual differences of electrons and neutrinos, like different masses and electric charges, are carried by the bound state \pref{expferm} \cite{Maas:2017wzi}. Only to leading order in $v$ in \pref{expferm} these properties are projected onto them, as shown in \pref{expferm2}.}.

Now, each of the individual cross sections violates the Bloch-Nordsieck theorem, which leads to a double-logarithmic Sudakov enhancement \cite{Ciafaloni:2000rp}. If $s\gg v$, they become
\begin{eqnarray}
\sigma_{\bar{l}^L l^L\to X}&=\sigma_{\bar{\nu}^L \nu^L\to X}=&A+S\\
\sigma_{\bar{l}^L \nu^L\to X}&=\sigma_{\bar{\nu}^L l^L\to X}=&A-S,
\end{eqnarray}
where $A$ contains the non-enhanced part, and $S$ the enhancement proportional to the exponentiated Sudakov logarithm. Inserting this into (\ref{tsx}) immediately shows that the enhancement cancels. The decisive step here is that it is necessary to do the summation, as mandated by gauge invariance. That they would cancel if summed follows from the BNT, and is well known \cite{Ciafaloni:2000rp}. Thus, the FMS mechanism explains how the summation can take place at high energies as mandated by gauge symmetry without altering the low-energy behavior accessed by LEP(2). Of course, this is now the behavior at small and large energies compared to the Higgs vacuum expectation value. What happens in the intermediate energy ranges where both behaviors transition into each other needs a full evaluation of \pref{me} in FMS-mechanism augmented perturbation theory, which is relegated to future work \cite{MPV:unpublished}.

However, to the current order, this exactly cancels the BNT violations. These have been estimated to be of the same order as the leading-order QCD effects at a 1 TeV lepton collider \cite{Ciafaloni:2000rp}, and substantially larger at a 3 TeV lepton collider \cite{Ciafaloni:2000rp,Bauer:2018xag}, depending on polarization. Having them removed by the BNT is therefore not a negligible effect.

Of course, eventually it will no longer be justified to treat the Higgs as spectators. If there are no non-trivial bound state contributions, this can be accounted for by augmented perturbation theory \cite{MPV:unpublished}. If this would not be the case, a PDF-type language \cite{Egger:2017tkd}, similar to the hadron case treated in section \ref{s:hadron}, would be necessary\footnote{See also \cite{Calmet:2000th,Calmet:2001rp,Calmet:2001yd} for a similar approach with a different motivation.}.

\section{Hadron scattering and PDFs}\label{s:hadron}

In principle, the same considerations apply as well with quarks in the initial states, i.\ e.\ for hadron colliders like the LHC or (future) hadron colliders \cite{Ciafaloni:2000rp,Fernbach:2020tpa}. In particular, the BNT is violated in the standard approach, and this has a substantial impact at sufficiently high energies  \cite{Ciafaloni:2000rp,Bauer:2018xag}. At first sight, it appears to be sufficient to follow the same prescription as before to fix it. But there is a subtlety, arising from the need to use PDFs to describe the hadron structure.

In the standard approach a cross section will be determined as \cite{Bohm:2001yx}
\be
\sigma_{PP\to X}=\sum_{ij}\int_0^1 dx\int_0^1 dy f_i(x) f_j(y)\sigma_{\bar{i}j\to X}(xp_1,yp_2),\label{pdf}
\ee
\no where $i$ runs over all constituents of both hadrons, the $f_i$ are the corresponding PDFs, and $X$ is the final state. Consider for the moment just the first generation of quarks. It is worthwhile to recall how BNT violations in the strong interactions are avoided. This happens as in the hard cross section an unbiased sum over all colors is performed \cite{Bohm:2001yx}. Each color appears equally as initial state in the hard cross section. This is due to the implicit presence of three PDFs for each quark, one for each color, which are identical\footnote{This is, of course, only possible in perturbation theory, as colored quarks are physical states in the usual BRST construction \cite{Bohm:2001yx}. However, the PDFs are genuine non-perturbative quantities, and the BRST construction is broken non-perturbatively \cite{Fujikawa:1982ss}, and in fact quarks as initial states cannot act as physical states \cite{Lavelle:1995ty}, and the PDFs need to be defined in a non-perturbatively manifest gauge-invariant way \cite{Ji:2013dva}. The reason why such a factorization nonetheless works is that in QCD there is a one-to-one correspondence between each quark and a gauge-invariant physical observable, its flavor. But it is also this feature, the flavor, which will be needed to be resolved at the electroweak level.}.

Because of the isospin sum rules
\bea
&&\int dx\left(f_{u^L}(x)-f_{\bar{u}^L}(x)+f_{u^R}(x)-f_{\bar{u}^R}(x)\right)=2\nn\\
\label{sr1}\\
&&\int dx\left(f_{d^L}(x)-f_{\bar{d}^L}(x)+f_{d^R}(x)-f_{\bar{d}^R}(x)\right)=1\nn\\
\label{sr2}\\
&&f_{i^L}(x)=f_{i^R}(x)\label{sr3}
\eea
\no the PDFs for the up quark and the down quark can never be the same. But since the left-handed up quarks and down quarks are the members of a weak multiplet similar to \pref{ldoublet}, this seems to imply that BNT violation cannot be avoided. Of course, a possibility appears to be to relax the equality of the left-handed quarks and right-handed quarks, and this is indeed anyhow necessary at high energies \cite{Bauer:2018xag,Manohar:2018kfx}. But the proton is a parity eigenstate, and thus fully shifting isospin to the right-handed quarks seems to be at least unlikely.

The reason for this impasse is, of course, that \pref{pdf} is really just the leading order of the FMS expansion. To understand this, it is helpful to start with the manifestly gauge-invariant description of the proton. To this end, it is useful to follow a two-step procedure \cite{Maas:2017wzi,Egger:2017tkd}. The left-handed quark doublet $q^L=(u^L,d^L)$ can be dressed in the same way as leptons in \pref{expferm}. This implies that the physical flavor of left-handed quarks is also nothing but the global SU(2)$_\text{c}$ symmetry of the Higgs field. Thus, left-handed quark flavor is the same for left-handed quarks and left-handed leptons. Their distinction is entirely by their other quantum numbers, like hypercharge, baryon number, and lepton number.

Construct then weakly-gauge-invariant Dirac spinors
\bea
U&=&\bma \left((X)^\dagger q^L\right)_1\cr u^R\ema\label{qcdquark}\\
D&=&\bma \left((X)^\dagger q^L\right)_2\cr d^R\ema\nn
\eea
\no which transform suitably under the diagonal $\mathrm{SU(2)_{f}}$ group. A nucleon operator can then be constructed as \cite{Egger:2017tkd,Maas:2017wzi,Gattringer:2010zz}
\bea
N&=&\frac{1}{2}\left(1+\gamma_0\right)\epsilon^{IJK}U_I\left(U^T_{J}C\gamma_5D_{K}\right)\label{nucleon}\\
&=&\frac{\epsilon^{IJK}}{2}\bma u^R-\left((X)^\dagger q^L\right)_1 \cr \left((X)^\dagger q^L\right)_1-u^R\ema_I\times\nn\\
&&\times\left(\left(\left((X)^\dagger q^L\right)_1^J\right)^T\tau^2\left((X)^\dagger q^L\right)_2^K-(u^R_J)^T\tau^2d^R_K\right)\nn\\
&=&\frac{\epsilon^{IJK}}{2}\left(\bma u^R-\phi_2u^L \cr \phi_2u^L-u^R\ema^I+\phi_1\bma d^L\cr -d^L\ema^I\right)\times\nn\\
&&\times\left(|\phi_2|^2(u^L_J)^T\tau_2d^L_K-|\phi_1|^2(d^L_J)^T\tau^2u^L_j-(u^R_J)^T\tau^2d^R_K\right.\nn\\
&&\left.+\phi_2\phi_1^*(u^L_J)^T\tau^2u^L_K-\phi_1\phi_2^*(d^L_J)^T\tau_2d^L_K\right).\nn
\eea
\no where $C$ is the charge conjugation matrix, $\tau^2$ is the second Pauli matrix, and the capital indices enumerate color. The projector in front is needed to project out positive parity, as needed for the proton.

In leading order in the Higgs vacuum expectation value, $\phi_2=v$ and $\phi_1=0$, \pref{nucleon} collapses to the usual expression in QCD \cite{Gattringer:2010zz}. The term in bracket acts as a diquark with zero net flavor. Spin and flavor is therefore carried entirely by the leading $U$, which mixes, due to the projector, left-handed components and right-handed components. It is thus visible how in a manifestly gauge-invariant description there are valence Higgs degrees of freedom in addition to valence quarks, and the former carry the flavor of the proton, very much in the same way as for leptons.

This now shows how \pref{pdf} can be seen as the leading order FMS expansion of the physical cross section. Every of the would-be quark PDFs describes the probability to encounter a flavor carrier in the proton, which are given by \pref{qcdquark}. Since the color indices play the same role as before, and at low enough energy the structure \pref{qcdquark} is not resolved, the hard cross section is then having Higgs-quark bound states as initial states. The cross section can then be FMS expanded like in \pref{loxs}. Keeping only the leading term yields finally the standard expression for hadronic cross sections \pref{pdf}.

If all internal weak structure of the proton \pref{nucleon} would be entirely contained in the  bound-state \pref{qcdquark}, it would be sufficient to switch from the leading-order FMS expansion to the full expression, like in \pref{tsx}, at high energies. This would maintain the BNT, at the expense of reinterpreting the quark PDFs as physical flavor PDFs.

If this is not the case, which may happen e.\ g.\ for multi-parton interactions, or if the separation between strong and weak bound state effects is no longer possible at high energies, this will require a different approach\footnote{At such energies the BEH effect becomes irrelevant, and thus factorization will work as in QCD, supplemented by a scalar. Literally the same steps will just need to be done, and the factorization scale is then given by the Higgs vev rather than the hadronic scale.}. In that case, it is necessary to introduce explicit Higgs PDFs\footnote{In \cite{Fernbach:2020tpa} this had been done only partially. With hindsight, this needs to be reinterpreted as attempting to encode the difference between the actual Higgs substructure and the conventional one only effectively by a single additional PDF, which did not influence the sum rule.} \cite{Fernbach:2020tpa} at the valence level to maintain the sum rules \prefr{sr1}{sr2}, as the flavor is carried by the custodial symmetry doublet eigenstate $(-\phi_1,\phi_2^*)$, which mixes the weak doublet eigenstate $(\phi_1,\phi_2)$. This can be accommodated by introducing four Higgs PDFs, $f_{hij}$, in which $i$ and $j$ denote the weak and custodial contributions, respectively. They therefore correspond directly to the entries of the matrix $X$ in \pref{eq:higgs_matrix} in terms of the required parton to participate in the hard interaction. This implies also
\be
f_{h\frac{1}{2}j}=f_{h-\frac{1}{2}j}\nn\\
\ee
\no to maintain gauge invariance and the BNT. Denoting for $j$ $u$ up-type flavor and $d$ for down-type flavor, this ultimately replaces \prefr{sr1}{sr2} by
\bea
\int dx\left(f_{u^R}(x)-f_{\bar{u}^R}(x)+f_{h\frac{1}{2}u}+f_{h-\frac{1}{2}u}\right)&=&2\nn\\
\int dx\left(f_{d^R}(x)-f_{\bar{d}^R}(x)+f_{h\frac{1}{2}d}+f_{h-\frac{1}{2}d}\right)&=&1\nn.
\eea
\no Of course, in $R_\xi$ gauges this appears to correspond to having unphysical states as possible initial states. However the Goldstone-Boson equivalence theorem relates them to longitudinally polarized vector boson matrix elements.

Note that in the baryon number sum rule still the left-handed quark PDFs appear, but they obey
\be
f_{u^L}=f_{d^L}\nn.
\ee
\no In the sumrule for momentum and electromagnetic charge now all PDFs, including the Higgs ones, appear.

This structure now automatically ensures that the BNT is also maintained in calculations of hadronic collisions. Of course, the different PDFs will need to be determined and evolved, including the other electroweak ones, along the lines of \cite{Bauer:2018xag,Manohar:2018kfx}. However, if at sufficiently low energies the weak substructure does not matter, it necessarily follows that
\bea
&&\sum_{u_L,d_L,hkl}\int_0^1 dx\int_0^1 dy f_i(x) f_j(y)\sigma^{\text{elementary}}_{\bar{i}j\to X}(xp_1,yp_2)\nn\\
&\approx&\sum_{u_L,d_L}\int_0^1 dx\int_0^1 dy \tilde{f}_i(x) \tilde{f}_j(y)\sigma_{\bar{i}j\to X}^{\text{elementary}}(xp_1,yp_2)\nn
\eea
\no need to hold, where the $\tilde{f}_i$ are now identical to the BNT-violating standard PDFs and the superscript ``elementary'' implies that these are the ordinary perturbative cross sections, i.\ e.\ the ones in leading order in the Higgs vev of FMS-augmented perturbation theory. However, since the factorization scale is crossed in this way, this is really not to be understood as an exact statement anymore. In fact, from the high energy perspective what happens is that either a valence left-handed quark from the structure \pref{qcdquark} or a left-handed quark radiated off from the valence Higgs takes part in the hard interaction, and the Higgs becomes off-shell suppressed. Neglecting the latter will then reinstantiate the BNT violation, but sufficiently small to be irrelevant at low energies.

Alternatively, if the weak substructure can be treated entirely using augmented perturbation theory, an approach like \pref{tsx} would also be possible. However, then a subtle problem arises, as this will not remove the strong substructure. Thus, there will be still subprocesses in which the QCD cores of the physical proton interact, which would need to revert again to a PDF language. But, as is visible in expression \pref{nucleon}, this process is {\it inclusive with respect to the left-handed weak doublets}. Thus the QCD expression involves at the QCD level not only proton-like states, but also other QCD-like objects. For these, PDFs are not as readily available. This may require other sources than experiments, e.\ g.\ using lattice calculations \cite{Ji:2013dva,Cichy:2018mum,Constantinou:2020hdm}, to follow such an approach instead of introducing Higgs PDFs.

In addition, similar considerations also apply to further generations. However, since off-diagonal intergeneration elements of the CKM matrix can be recast as intergeneration Yukawa interactions of the Higgs, it may in these cases probably be better to set up the PDF scheme in such a basis. However, this is beyond the scope of this work, as the BNT restoration will work in the same way also for the other generations.

\section{Beyond the Bloch-Nordsieck theorem}\label{s:kln}

It has been established early on in QCD that it is insufficient at higher orders to just satisfy the BNT to avoid the appearance of infrared enhancements \cite{Agarwal:2021ais,Doria:1980ak,Andrasi:1980qw}. Rather, satisfying the KLNT \cite{Kinoshita:1962ur,Lee:1964is} would be necessary \cite{Doria:1980ak,Andrasi:1980qw,Agarwal:2021ais}. This is a non-trivial task, in view of perturbatively massless gluons, as it requires summing over all energy-degenerate states in both the initial state and the final state. As the electroweak case at high energies is really just scalar QCD, it has to be expected that the same problem occurs also in the present case \cite{Ciafaloni:2022kil}.

This issue can be resolved in QCD using a suitable factorization \cite{Agarwal:2021ais} approach. It stands to reason that this is, at least formally, analogously applicable to the present case. But since the KLNT requires a summation in both the initial state and the final state, this necessitates first introducing the equivalent of fragmentation and hadronization in the electroweak case. Since already in QCD, this becomes a complex, non-perturbative process, it is beyond the scope of the present work.

There is, however, one possible saving grace. In contrast to the QCD case, there are in fact not infinitely many energy-degenerate initial states and final states. While the masses of the gauge bosons and the Higgs have been neglected, they are actually non-zero. This manifests itself in the appearance of the $W$ mass rather than of the vev $v$ in the Sudakov logarithm. Thus, the number of energy-degenerate states turns out to be small, and not infinite as in the QCD case, provided the masses are explicitly accounted for. This could allow to satisfy the KLNT in the very same way as the BNT in the present work, and requires only the inclusion of the lightest complete gauge multiplet. However, this is again beyond the scope of this work, especially as the mass corrections are far form trivial to include in such calculations \cite{MPV:unpublished}.

\section{Conclusion}\label{s:sum}

Manifest gauge invariance enforces the use of composite operators to describe physical states also for the weak interactions. However, the combination of the BEH effect and the FMS mechanism shows that at sufficiently small energies in the standard model this does yield quantitatively essentially the same results as if only perturbative gauge invariance is respected. Serious problems only arise when kinematics are no longer determined primarily by the FMS-dominant constituent, i.\ e.\ in most cases at either extremely small \cite{Jenny:2022atm,Maas:2018ska} or extremely large energy transfers \cite{Maas:2020kda,Egger:2017tkd,Fernbach:2020tpa,Dudal:2020uwb}. It then becomes necessary to revert to a full desciption of the composite states.

We have investigated how this could restore the BNT by using manifestly weakly gauge-invariant initial states, at least at leading order. This potentially affects cross sections at very high energies considerably, especially at future TeV-scale lepton colliders. The situation is the same at hadron colliders, but here the PDF structure both suppresses the effect \cite{Ciafaloni:2000rp} and also makes its resolution more involved. Likewise, we expect final states, including fragmentation, to require a similar treatment to avoid violations of the KLNT, which is beyond the scope of this work.

Augmenting perturbation theory with the FMS mechanism may allow for a possibility to take these effects into account with only marginally more effort than in usual perturbation theory. Given the potential impacts for predictions for future colliders, this appears a small price to pay.

\begin{acknowledgments}
We are grateful to S.\ Pl\"atzer and R.\ Sondenheimer for helpful discussions, to S.\ Pl\"atzer for a critical reading of the manuscript, and to B.\ Webber for prompting this research. This work was done within the scope of the FCC Feasibility Study.
\end{acknowledgments}

\bibliographystyle{bibstyle}
\bibliography{bib}

\end{document}